# Role of substrate temperature at graphene synthesis in an arc discharge

Xiuqi Fang, [1,a] Alexey Shashurin,[1] and Michael Keidar[1,a]

*Department of Mechanical & Aerospace Engineering,*

*The George Washington University, Washington DC, 20052, USA*

*School of Aeronautics and Astronautics,*

*Purdue University, West Lafayette, Indiana, 47907, USA*

**Abstract**

The substrate temperature required for synthesis of graphene in an arc discharge plasma was studied. It was shown that an increase of copper substrate temperature up to the melting point leads to an increase in the amount of graphene production and the quality of graphene sheets. Favorable range of substrate temperatures for arc-based graphene synthesis was determined, and it is in a relatively narrow range of about 1210-1340 K.

1. **INTRODUCTION**

Graphene is a flat, one-atom-thick closely packed honeycomb lattice sheet, which consists of sp2-bonded carbon atoms.[1,2] This material combines aspects of semiconductors and metals.[3] Graphene is currently being studied for a wide range of applications such as transparent electrodes,[4] graphene field-effect transistors,[5,6] paper-based ultra-capacitors,[7,8] tissue enhancement additives[9] and biochemical sensors.[10]

Several approaches have been used to synthesize graphene flakes, including mechanical and chemical exfoliation,[11] chemical vapor deposition (CVD),[12] and graphene growth on insulating silicon carbide (SiC).[13] Generally, these methods allow the synthesis of two types of graphene, namely large-area pristine graphene films and graphene nanoplatelets in bulk quantities.[14,15,16] Many of these synthesis techniques are accompanied with certain disadvantages that require research on improving the synthesis approaches. Mechanical exfoliation is very expensive and an ineffective method characterized by low output.[1,17] Chemical exfoliation could generate by-products of toxic gases or hazardous chemical components during the synthesis process.[18] The main challenges in the CVD process are the production of large-area, high quality continuous and uniform graphene films, and the reduction of damages to the graphene film during transfer.[19,20,21] Current production capability of bulk graphene nanoplatelets by chemical exfoliation are quite limited and estimated to be around several hundred tons of material annually worldwide. [17,22,23] The source of carbon could even be achieved from some raw carbon-containing material instead of expensive purified chemicals.[24]

Another method to synthesize bulk graphene nanoplatelets is arc discharge. It has been shown to successfully synthesize graphene flakes with and without the enhancement of a magnetic field.[25,26,27,28,29,30,31] Graphene was synthesized using anodic arc ablation in a helium environment at atmospheric pressure. Samples were collected from the chamber wall or the surface of the magnet, and were characterized with SEM, TEM, AFM and RAMAN spectroscope. Graphene production with the arc discharge method is a promising technique, as it is environmentally friendly and yields high purity graphene flakes compared to chemical exfoliation.[32] Plasmas of the arc discharge can also provide a convenient way to add functional groups in situ during synthesis. Recently this has also been demonstrated by plasmas of glow discharge in a current experiment.[33]

However, the specific range of temperatures required for graphene synthesis in the arc discharge have not been investigated. In this paper, ablation of carbon electrodes by means of an arc discharge in a helium atmosphere and the following delivery to the heated substrate has been studied. The work is focusing on exploring the correlation between the synthesis substrate temperature and the properties of arc-produced graphene. Due to the difficulties in measuring temperature distribution along the small substrate, temperature simulations were employed to determine the substrate temperature at the synthesis location instead.

## 2. METHODS

### 2.1. Experimental set up and synthesis procedure

The synthesis setup consists of a stainless steel cylindrical vacuum chamber with a total volume of 4500cm$^3$ (27 cm in length and 14.5 cm in diameter). A pair of electrodes, a cathode and anode, is installed along the vertical axis of the chamber. Both electrodes are made of POCO EDM-3 graphite. The cathode is a cylindrical rod with a diameter of 13 mm, while the anode is a hollow tube with inner and outer diameters of 3 and 5 mm respectively.

Figure 1a shows the schematic of the arc discharge synthesis setup. The anode and cathode were placed 2 mm apart, and the distance between the substrate and the cathode-anode assembly was about 1.5-2 cm. A shutter (shown in Figure 1b) made of molybdenum foil, with a quarter section cutout, allowed control of the substrate exposure time to the arc plasmas. The vacuum chamber was pumped to the pressure of about 13 Pa and then high purity helium (about 99.97%) was introduced into the chamber to the pressure around 67000 Pa.[30]

A 1.5 x 14 mm$^2$, 0.1 mm thick copper foil was used as a substrate to collect the graphene sample. In order to heat the substrate, a nickel-chromium resistance wire was used. The resistance wire was wound into the shape of a spring, and the copper foil was then inserted inside. Voltage generated by a variable auto-transformer was applied to the heater during the synthesis. Before the experiment, the substrate was pre-cleaned using ethanol. It has to be noted that extreme requirements to residual vacuum and substrate cleaning were not applied in this work since bulk graphene nanoplatelets synthesis was pursued rather than pristine graphene.

The arc electrodes were connected to an external DC power source at a fixed arc current of about 75A. Arc current and arc voltage were recorded by a digital oscilloscope. The anode motion was controlled by a linear drive system using a personal computer equipped with a National Instruments Data Acquisition card, and the program was written on the LABVIEW platform.[34] The arc was generated by mechanical contact of the arc electrodes followed by their immediate separation. A camera was utilized to record the whole process of the synthesis procedure.

The graphene synthesized on the copper substrate was then characterized using a Horiba LabRAM spectroscope, SIGMA VP-02-44 SEM and JEOL 1200 EX TEM. It has to be noted that there was no graphene observed on the pre-heated substrate if the arc discharge was not initiated, indicating that possible production of graphene as a result of decomposition of hydrocarbon oil utilized in the vacuum pumping system is negligible.

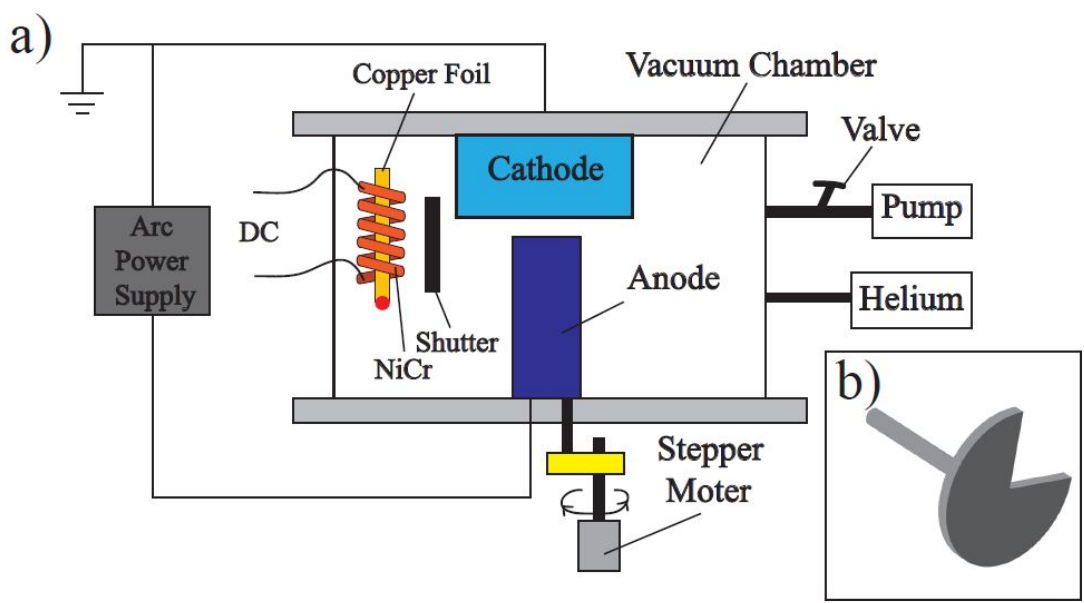

**Figure 1: a) Schematic of the set-up of the plasma based synthesis system**
**b) Shutter used between the substrate and arc**

The substrate was heated for 2 minutes before initiating the arc. Once the arc stabilized, a few seconds after initiation, the shutter then made one full revolution for one second. After a single revolution, the arc was then turned off. Heating of the copper foil was continued for an additional 3 minutes and then was also turned off.

## 2.2. Numerical simulations

Numerical simulations were conducted in order to determine the temperature distribution along the copper substrate, which was utilized in the experiment. Numerical simulations were carried out by COMSOL 5.0 software.

A thermal model of the copper substrate in the helium volume utilizes the heat transfer equation shown below. The heat transfer equation takes into account the heat conduction in both the copper substrate and helium, heat radiation from the substrate surface and heat convection:

$$\rho C_P \frac{\partial T}{\partial t} = \nabla \cdot (k \nabla T) \qquad (1)$$

where $\rho$ – copper/helium density, $C_p$ – heat capacity at constant pressure for copper/helium, $T$ – temperature, $k$ – thermal conductivity of copper/helium. The steady-state solution ($\frac{\partial T}{\partial t} = 0$) of this thermal problem was obtained.

The boundary conditions were formulated as follows. Temperature of one end of the foil piece was set to be equal to copper's melting temperature in order to model the experimental fact that one end of the foil was observed to be melted. All surfaces of the foil were exchanging heat with their surroundings by means of heat convection and Stefan-Boltzmann radiation as follows:

$$Q_c = h \cdot (T_{ext} - T) \qquad (2)$$

$$Q_r = \varepsilon \sigma (T^4_{amb} - T^4) \qquad (3)$$

where $T_{amb}$ – ambient temperature, $\sigma$ – Stefan-Boltzmann constant of copper, $h$ – heat transfer coefficient, $T_{ext}$ – external temperature and $\varepsilon$ – emissivity of copper. $T_{ext} = T_{amb} = 300$ K were chosen in simulations.[35]

It has to be noted that precise conditions of the heat exchange between the substrate and the surroundings are unknown primarily due to the large spread of the heat transfer coefficient $h$ and copper emissivity $\varepsilon$. Thus we conducted temperature simulations for the extreme range of these parameters, namely ε = 0.05 - 0.87 and $h$ = 4.33 - 80 W/(m²K).[36] The extreme values, ε = 0.87 and $h$ = 80 W/(m²K), providing the maximal temperature drop along the copper substrate, were used in the results presented below. The physical properties of copper utilized in the simulations are summarized in Table 2.

| Property | Value |
|---|---|
| Thermal Conductivity, k [W/(m K)] | 400 |
| Density, ρ [kg/m³] | 8700 |
| Heat Capacity, $C_p$ [J/(K kg)] | 385 |
| Emissivity, ε | 0.87 |
| Heat transfer coefficient, h [W/(m² K)] | 80 |

Table 1: Physical properties of copper utilized in simulations

## 3. RESULTS AND DISCUSSIONS

The calculated steady-state temperature distributions along the substrate are shown in Figure 2. One can see that the temperature of the melting point of the copper substrate is present on one end (which is 1358 K), and the temperature gradually decreases along the copper substrate. Temperature distribution in the surrounding helium is also shown.

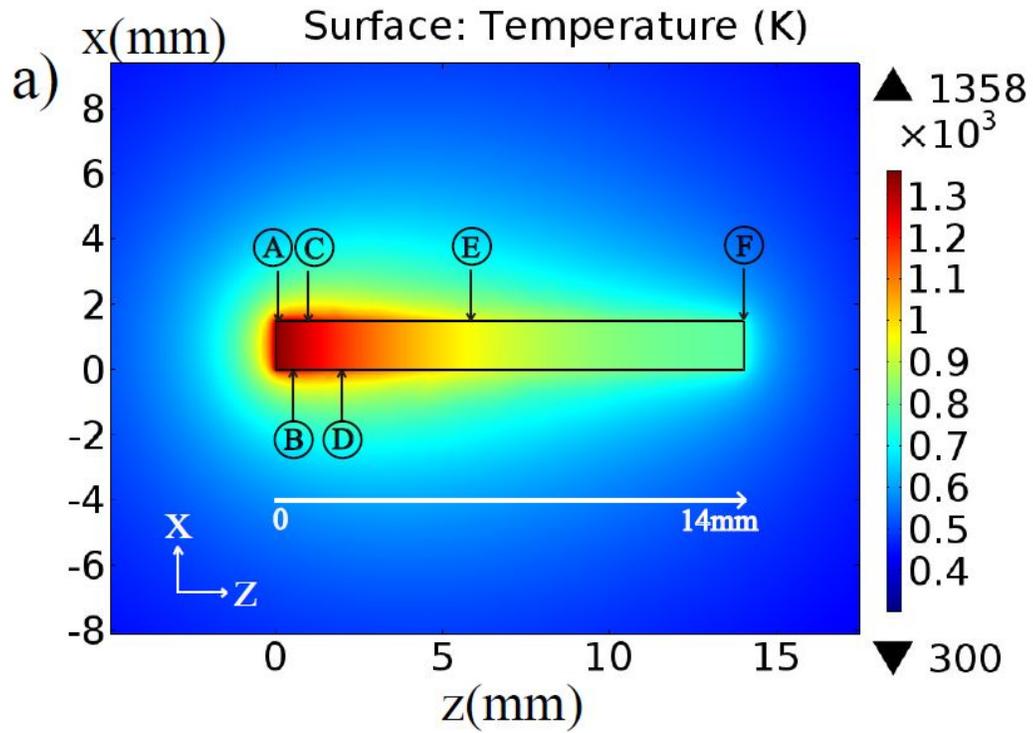

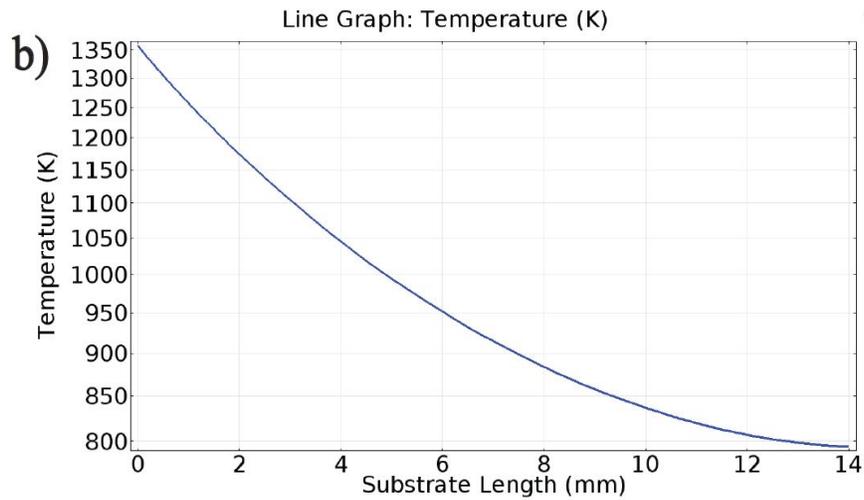

**Figure 2. a) The calculated 2D steady-state temperature distributions of the copper substrate surface. Six points A - F (A = 0mm B = 0.5mm C = 1mm D = 2mm E = 6mm F = 14mm) were chosen along the copper substrate and further characterization such as RAMAN spectroscopy and SEM were performed on those points**

**b) Temperature distribution along z-axis**

Figure 3 shows RAMAN spectrograms taken at points (A) - (F) [see Figure 2a] along the substrate. Three intense bands typical for graphene, namely D, G and G' were observed along the spectrogram at around 1360 cm$^{-1}$, 1565cm$^{-1}$ and 2700 cm$^{-1}$, respectively (see Figure 3a). The D-band is associated with the amount of defects in graphene sp2 bonds. G-band and G' are typical for all sp2 carbons.[37,38,39,40] The ratio of band intensities I(D)/I(G) could be used to characterize the level of disorder in graphene.[41] G band splits into G band (1600 cm$^{-1}$) and D' band (1620 cm$^{-1}$) if there are any distributed impurities or surface changes in the graphene.[42,43] In this experiment, D'-band is caused by the oxygen impurities on the sample.[44] G* band (around 2450 cm$^{-1}$) can be assigned as an overtone mode of LO phonon. G' band is the two-phonon band, which is allowed in the second order RAMAN spectrogram of graphene.[38] The ratio of G'-band and G-band (I(G')/I(G)) could be used to estimate the number of layers in the graphene flakes and differentiate between the graphene and the amorphous carbon.[45] Table 2 shows the ratio I(G')/I(G) for the locations (A) - (F) and the temperature at the corresponding location. Using a 632.8nm excitation laser, the full width at half maximum (FWHM) of G and G' band for monolayer graphene is 15cm$^{-1}$ and 24cm$^{-1}$.[46] For multilayer graphene the FWHM for G' band is around 60cm$^{-1}$.[38] In addition, all bands' FWHM broaden due to disorder.[47] From Figure 3a, the FWHM for G band is about 20cm$^{-1}$ and for G' band is about 60cm$^{-1}$. Based on these results it might be concluded that the sample synthesized in arc discharge method represents several layer graphene. Figure 4 shows SEM images taken at locations (A) - (F). Figure 5a shows the TEM images of the graphene sample using mechanical ways to transfer from the copper surface to the TEM grid. This TEM image shows the typical flake structures for several layer graphene. And it was taken with a Joel-1200 TEM machine and it was captured by the probe with a 100V bias. Figure 5b shows the diffraction pattern which is arrayed in a hexagonal manner, and dots are visible. These dots present evidence of well-ordered thin sheet of single crystal structures.[48,49]

RAMAN spectrograms and SEM images taken at points (A) through (D) indicate the presence of large amounts of square-shaped multi-layered graphene flakes attached to the surface of the copper foil around the melted side of the sample. The ratio of I(G')/I(G) decreased from about 0.5 at the melted sample edge to about 0.25 at 2 mm from the edge. At the same time, visual observation of the copper foil did not demonstrate any darkening of the normal color of the foil due to the graphene deposition, which indicates that the deposition was relatively thin and nearly transparent. Locations further away ($T$<1210 K) from the melted zone of the sample [see points (E) and (F)] indicate the absence of G'-band and demonstrate RAMAN spectrograms typical for amorphous carbon structures.[45]

Based on the RAMAN spectrogram and the fact that temperature simulations were conducted at conditions providing maximal temperature drop along the sample (see above), it may be concluded that the substrate temperatures around the melting point of copper (1210-1340K) are required for the graphene synthesis. It has to be noted that these synthesis temperatures in arc-based graphene synthesis are higher than that for PECVD processes (about 752K).[50]

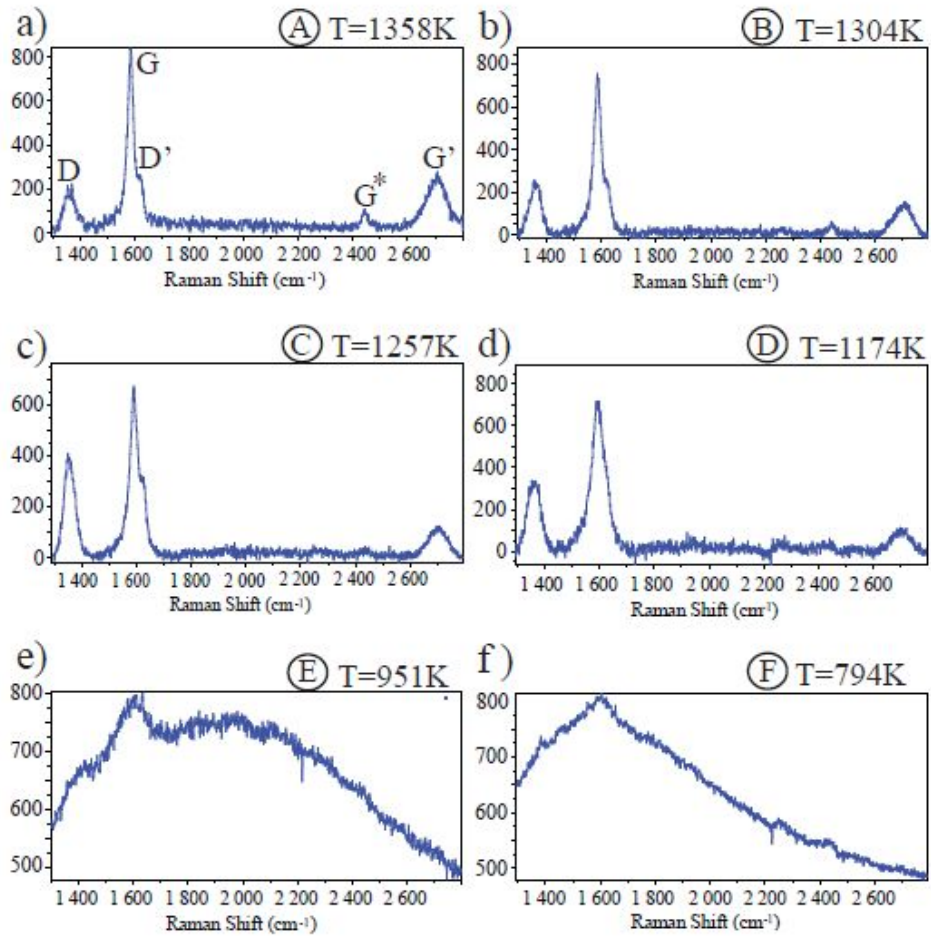

Figure 3. RAMAN spectrogram for the 6 points chosen along the substrate at their estimate

temperature

| Point | z/mm | T/K | I(G')/I(G) |
|-------|------|------|------------|
| A | 0 | 1358 | 0.5 |
| B | 0.5 | 1304 | 0.4 |
| C | 1 | 1257 | 0.25 |
| D | 2 | 1174 | 0.15 |
| E | 6 | 951 | 0 |
| F | 14 | 794 | 0 |

**Table 1: I(G')/I(G) ratio related to z and T**

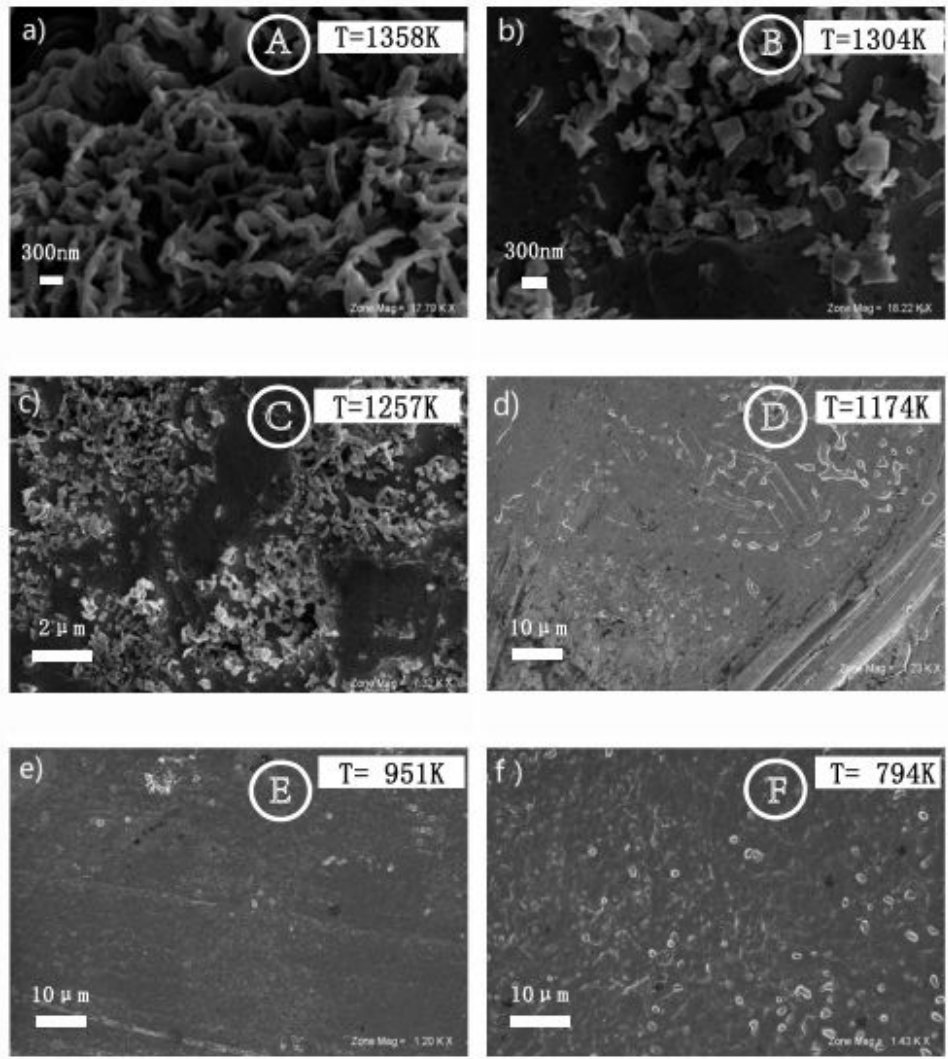

**Figure 4: SEM images for the 6 points chosen along the arc produced graphene sample substrate with their estimated temperature.**

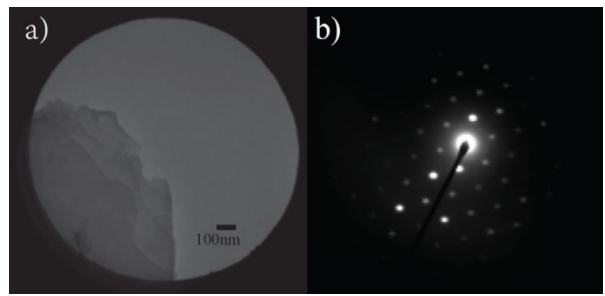

**Figure 5: TEM images of graphene flakes synthesized in arc discharge. The insert image shows the electron diffraction pattern of the graphene flakes in the selected area.**

SEM images of the synthesized graphene indicate flake attachment to the copper substrate over at least part of the flake surface (see Fig 6, taken high magnification at point A). This may indicate that the graphene film was first synthesized fully attached to the substrate surface, and then it was partially detached from the substrate and torn apart during the substrate cool down. Since copper is characterized by very low carbon solubility,[51] it may be hypothesized that the surface-catalyzed process involving diffusion of carbon species along the substrate surface, followed by its attachment to the edge existing graphene domains, was employed at synthesis. Relatively high temperatures very close to the copper substrate melting point, which were observed in the experiments, were necessary to produce the graphene nanoplatelets. It can possibly indicate that a relatively high mobility of carbon species along the surface is required for the synthesis.

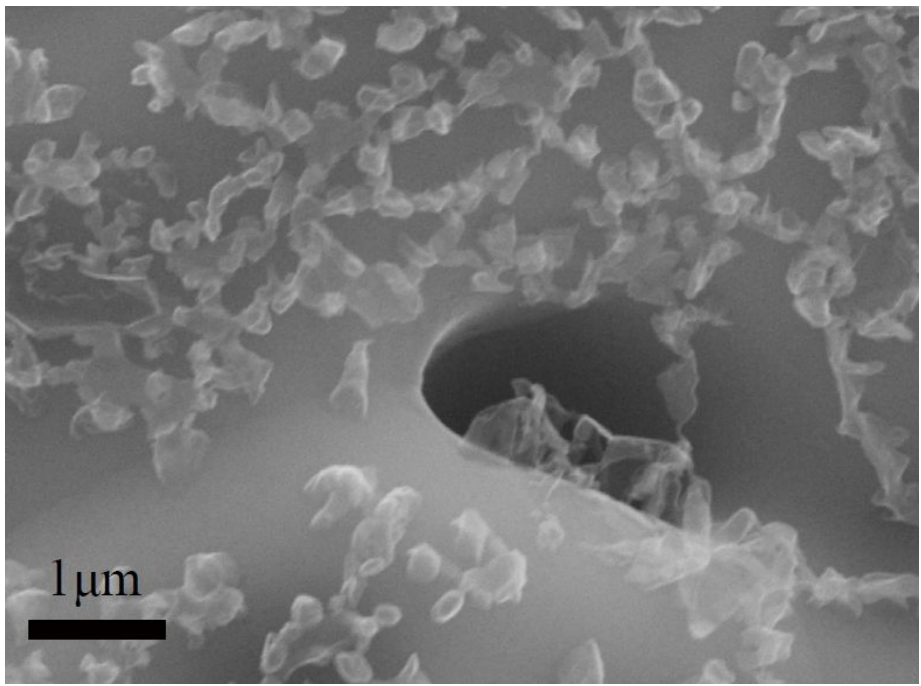

**Figure 6. SEM image taken at high magnification at point A**

**4. CONCLUSION**

The arc discharge technique was utilized for graphene synthesis on copper foil substrates. It has been shown that temperatures very close to the substrate melting point (within 150K) are required to produce the graphene nanoplatelets in an arc discharge. In contrast, lower temperatures result in the production of the amorphous carbon structures.

**ACKNOWLEDGEMENTS**

This work was supported by the U.S. Department of Energy, Office of Science, Basic Energy Sciences, Materials Sciences and Engineering Division.